Supplementary Information for:

# Key role of the Madden-Julian Oscillation on tropical and subtropical humid heat and heatwaves

Claire Rocuet,[a,b] Takeshi Izumo,[b,c] Bastien Pagli,[a,b,c] Neil J. Holbrook,[d,e]

Sophie Cravatte,[c,f] Marania Hopuare,[a,g] Maxime Colin[h]

[a] *Université de la Polynésie Française (UPF), Tahiti, French Polynesia*

[b] *UMR241 SECOPOL laboratory (IRD/UPF/IFREMER/ILM), Tahiti, French Polynesia*

[c] *Institut de Recherche pour le Développement (IRD), France*

[d] *Institute for Marine and Antarctic Studies, University of Tasmania, Hobart, Tasmania, Australia*

[e] *Australian Research Council Centre of Excellence for the Weather of the 21st Century, University of Tasmania, Hobart, Tasmania, Australia*

[f] *Université de Toulouse, LEGOS (CNES/CNRS/IRD/UT3), Nouméa, Nouvelle-Calédonie*

[g] *GePaSud Laboratory, Tahiti, French Polynesia*

[h] *Leibniz Centre for Tropical Marine Research, ZMT, Bremen, Germany*

*Corresponding author*: Claire Rocuet, claire.rocuet@doctorant.upf.pf

## 1. Introduction

The present supplementary information includes a table, steps followed to calculate heat stress indices and surface equivalent potential temperature and 17 figures to support the main text of the manuscript.

| Phases | November to April (NDJFMA) | May to October (MJJASO) |
|---|---|---|
| 1 | n = 522 | n = 408 |
| 2 | n = 575 | n = 702 |
| 3 | n = 605 | n = 663 |
| 4 | n = 594 | n = 305 |
| 5 | n = 496 | n = 455 |
| 6 | n = 583 | n = 658 |
| 7 | n = 598 | n = 580 |
| 8 | n = 577 | n = 434 |

Table 1. Number of days (n) for each phase during the two seasons NDJFMA and MJJASO when the MJO is active (OMI amplitude >1) between May 1979 and October 2020.

## 2. Calculation of OMI

The OMI dataset used here is provided by the NOAA PSL, Boulder, Colorado, USA, via their website: https://psl.noaa.gov/mjo/mjoindex/. The OMI involves an EOF analysis of 30–96-day eastward only filtered daily OLR data between 20°N and 20°S using a 121-day sliding window centred on each day of the year to capture the seasonal shifts of the MJO from 1979-2012 (Kiladis et al., 2014). The 20-96 day filtered daily OLR data from 1979 to 2021, including both eastward and westward wavenumbers (up to zonal wavenumber 72), are then projected onto the corresponding spatial EOFs associated with that day of the year using PC analysis, reducing the data to the first two principal components (PC1 and PC2), which effectively represent the MJO and BSISO's variability. This approach ensures the OMI's adaptability to seasonal changes and provides a robust measure of the MJO's phase and strength. To align the OMI with the RMM convention for comparison, the sign of the OMI PC1 is reversed, and the principal component ordering is swapped, making OMI(PC2) equivalent to RMM1 and -OMI(PC1) equivalent to RMM2. The combination of PC1 and PC2 produces a two-dimensional phase space divided into 8 sections, representing the 8 phases of the MJO. It is used to measure the location and amplitude of the active convective pattern of the MJO.

## 3. Calculation of Heat Index

The following method outlines the steps for calculating the Heat Index (HI) developed by Rothfusz (2019) based on Steadman's calculations (Steadman, 1979). The following equations are in degrees Celsius (°C), based on formulas (originally in °F) provided by the NOAA online:
https://www.wpc.ncep.noaa.gov/html/heatindex_equation.shtml

Firstly, we calculate the surface relative humidity ($RH$) in %:

$$RH_{2m} = \frac{e(T_{2m})}{e_s(T_{2m})} * 100 = \frac{e_s(T_D)}{e_s(T_{2m})} * 100 \quad (1)$$

where $e$ is the water vapor pressure in mb and $e_s$ is the saturated water vapor pressure in mb. $T_{2m}$ and $T_D$ are respectively the 2-m dry-bulb temperature and the 2-m dew point temperature expressed in Kelvin. $e_s(T)$ is obtained thanks to the August–Roche Magnus formulation of the Clausius–Clapeyron equation:

$$e_s(T) = 6.1094exp\left(\frac{a(T-C)}{T-C+b}\right) \quad (2)$$

where a=17.625 and b=243.04 and C= 273.15 K are constants. $e_s(T_{2m})$ and $e_s(T_D)$ are computed with this equation. In the following equations, $T$ and $RH$ correspond to $T_{2m}$ and $RH_{2m}$.

1. Initial Linear Formula:

$$HI_0 = 1.1T + 0.026RH - 3.944 \tag{3}$$

2. Refinement Step:

$$HI = \frac{HI_0 + T}{2} \tag{4}$$

3. If $HI > 26.667°C$ we use the quadratic formula (5), else keep HI computed with (4) :

$$HI = c_1 + c_2T + c_3RH + c_4RHT + c_5T^2 + c_6RH^2 + c_7RHT^2 + c_8RH^2T + c_9RH^2T^2 + ADJ \tag{5}$$

where:

$c_1 = -8.78469476,$

$c_2 = 1.61139411,$

$c_3 = \ 2.33854884,$

$c_4 = -0.14611605,$

$c_5 = -1.230809 \times 10^2,$

$c_6 = -1.642483 \times 10^2,$

$c_7 = 2.21173 \times 10^3,$

$c_8 = 7.2546 \times 10^{-4},$

$c_9 = -3.58 \times 10^{-6},$

And the adjustments *ADJ*:

$$ADJ = \{-\frac{13 - RH}{7.2}\sqrt{\frac{17 - |1.8T - 63|}{17}}\ if\ 26.667 < T < 44.444\ °C\ and\ RH < 13\% \quad -\frac{RH - 85}{18} . \frac{55 - 1.8T}{5}\ \ if\ 26.667 < T < 30.556\ °C\ and\ RH > 85\%\ \ 0\ else$$

## 4. Calculation of simplified Wet Bulb Globe Temperature

The simplified wet bulb globe temperature, or sWBGT, is an approximation of the WBGT originally designed for estimating heat stress in sports medicine and adopted by the Australian Bureau of Meteorology (American College of Sports Medicine, 1984; Australian Bureau of Meteorology). This approximation ignores the variations in the solar radiation or windspeed, as it assumes a moderately high radiation level and light wind conditions. Although its accuracy in representing the original WBGT index may be questionable, we chose to use it for comparison to the Heat Index due its wide use. The simplified formula is:

$$sWBGT = 0.567T + 0.393e + 3.94 \tag{6}$$

where $T$ is the is 2m temperature in °C and $e$ is the water vapor pressure (in mb):

$$e = e_s(T)\frac{RH}{100} \tag{7}$$

## 5. Band-pass filter and composite analysis

ERA5 hourly data were initially interpolated to a 1° x 1° longitude-latitude grid and aggregated into daily averages (i.e. same spatiotemporal resolution as GPCP). The daily climatology (Fig. S2) and anomalies were computed for the entire available time periods of the ERA5, NOAA OLR and GPCP datasets. A 20–96-day passband filter was applied to the daily anomalies using the difference of two lowpass Lanczos filters (window length of 192 days, half-power frequencies of 1/20 $day^{-1}$ for the high frequency filter and 1/96 $day^{-1}$ for the lower frequency filter). Composites were generated by averaging the anomaly field data over all the days falling within each MJO phase, and with MJO amplitude>1, using the OMI index for a particular season. To evaluate if the composites of the field anomalies significantly differ from zero at the 95% confidence level, statistical significance was assessed using a two-tailed student *t*-test. The number of effective degrees of freedom associated with each phase was calculated based on the discontinuous time-series, assuming an effective degree of freedom every 5 days (an approximation based on Equation 30 of Bretherton et al. (1999)). The number of days for each phase for each season for the 1979-2020 period is presented in Table S1. Tw anomalies for each phase within the given season are normalised using the standard deviation (STD) of the season's unfiltered Tw daily anomalies. This effectively removes the seasonal cycle and filters out components of interannual and long-term variability (Fig S3a,d). Tw anomalies normalised by the STD of intraseasonal (20–96 day) Tw anomalies are shown in the SI (Fig. S3a,c, Fig. S4).

## 6. Supplementary Figures

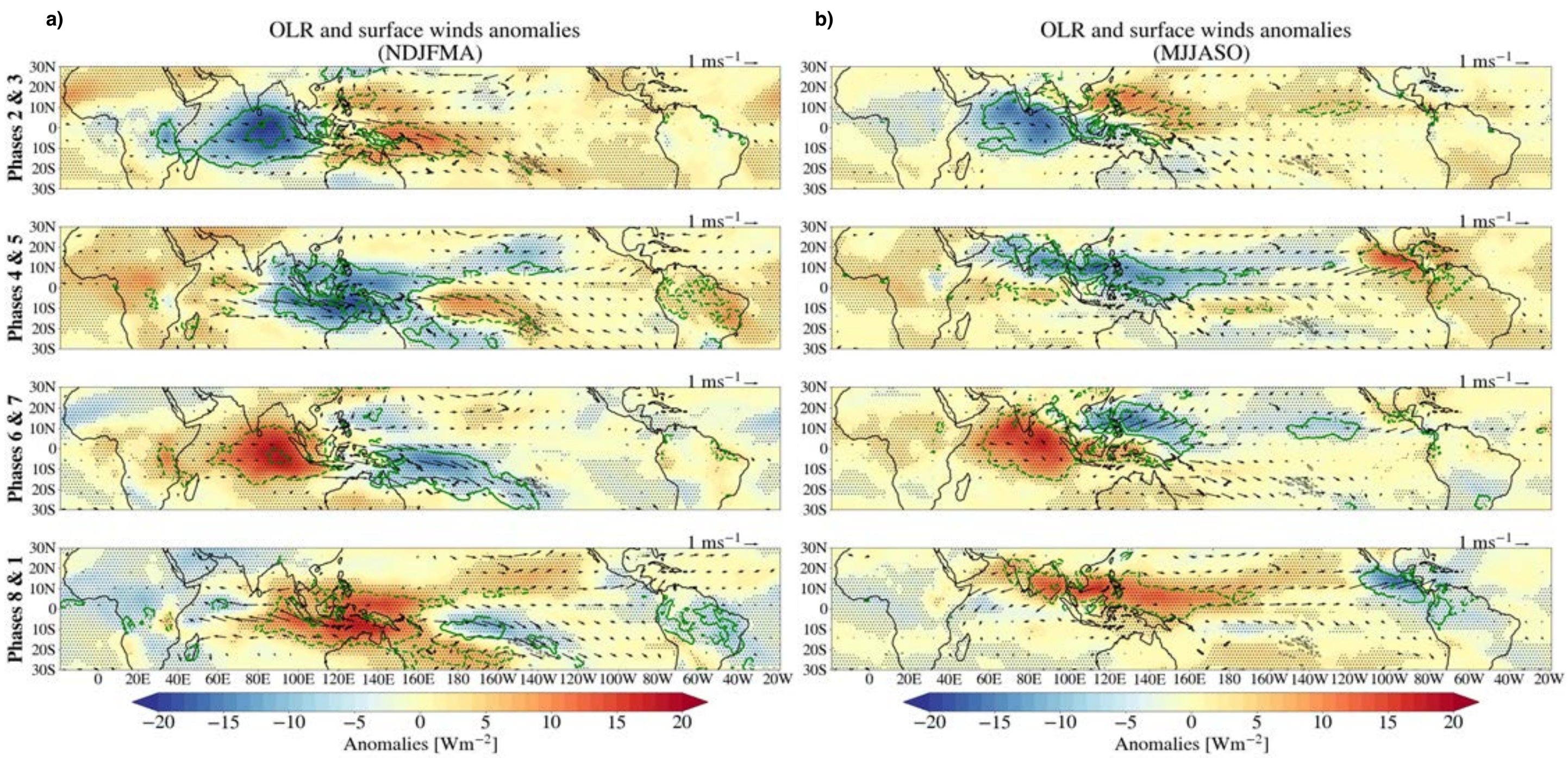

Fig. S1. As main Fig. 1 but using the RMM MJO index instead of the OMI for comparison. The RMM index dataset is provided by the BOM and is available online: http://www.bom.gov.au/climate/mjo/graphics/rmm.74toRealtime.txt.

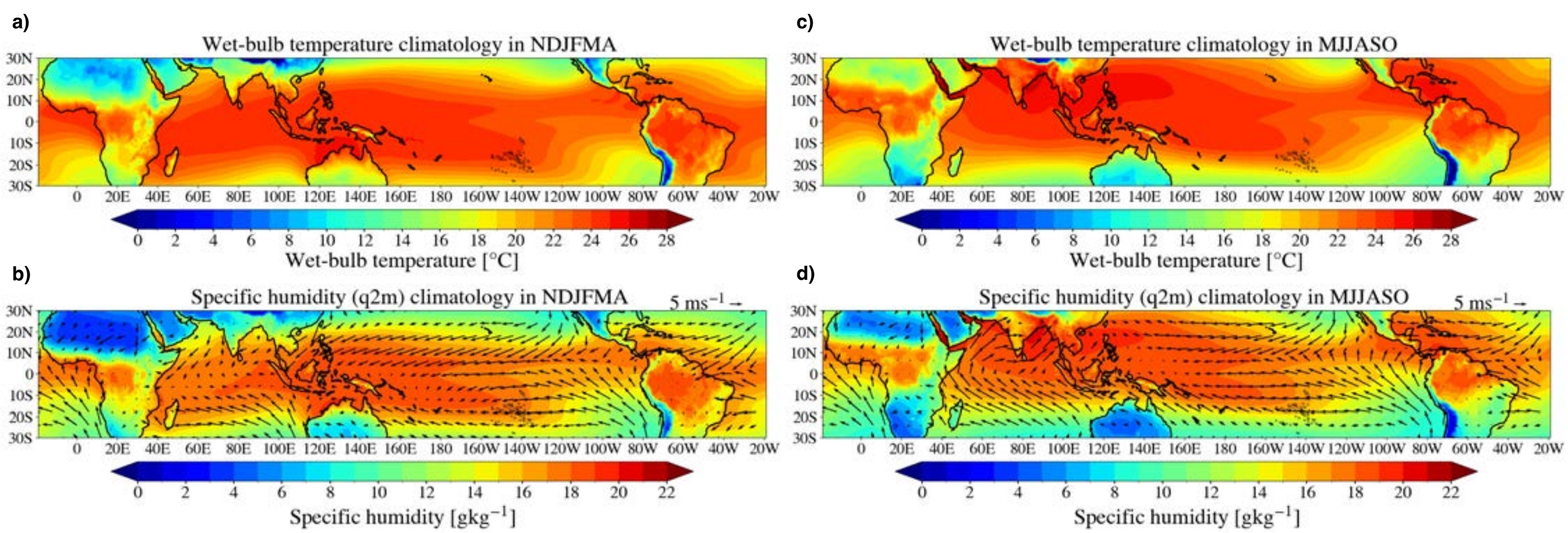

Fig. S2. Wet-bulb temperature climatology (°C, colour shading) and specific humidity (q2m) (colour shading) with 10m wind climatology (vectors) in a-b) NDJFMA and c-d) MJJASO.

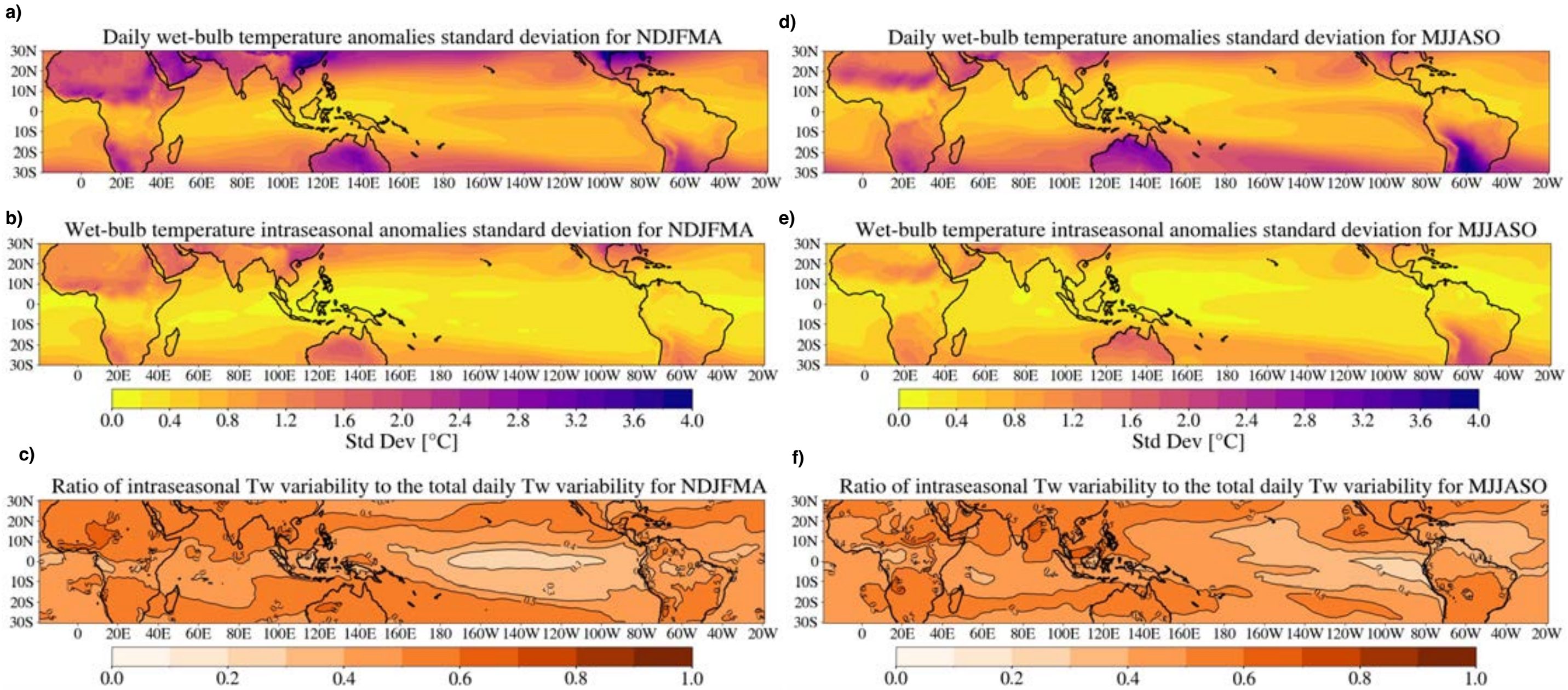

Fig. S3. Standard deviation (STD) of the unfiltered Tw anomalies (i.e., total variability without the seasonal cycle), the STD of the intraseasonal Tw anomalies and their ratio in a-c) NDJFMA and d-f) MJJASO.

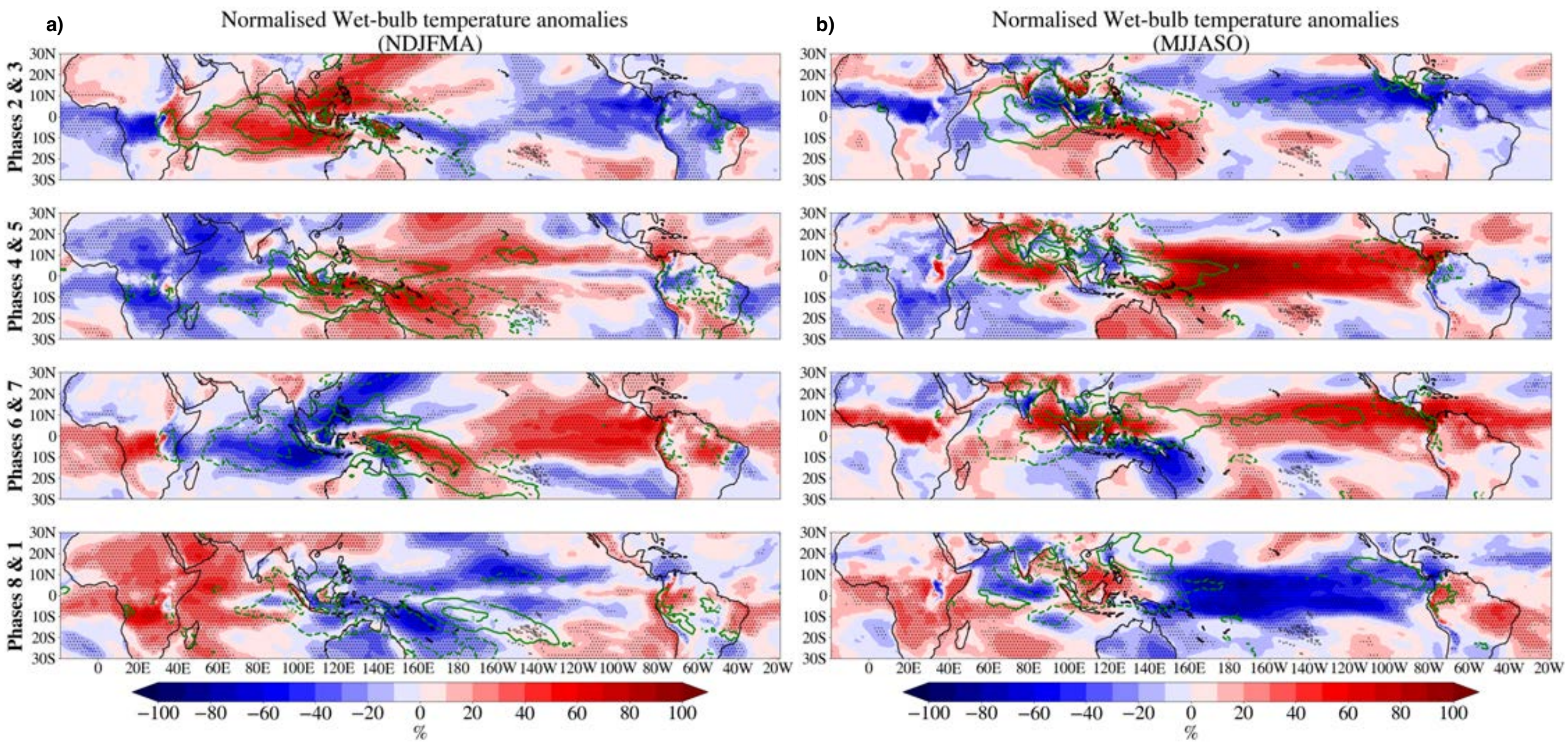

Fig. S4. Same as Fig. 2 but expressed as a percentage of the standard deviation (STD) of the intraseasonally filtered Tw anomalies (i.e., 20-96 days variability), with an adjusted color scale.

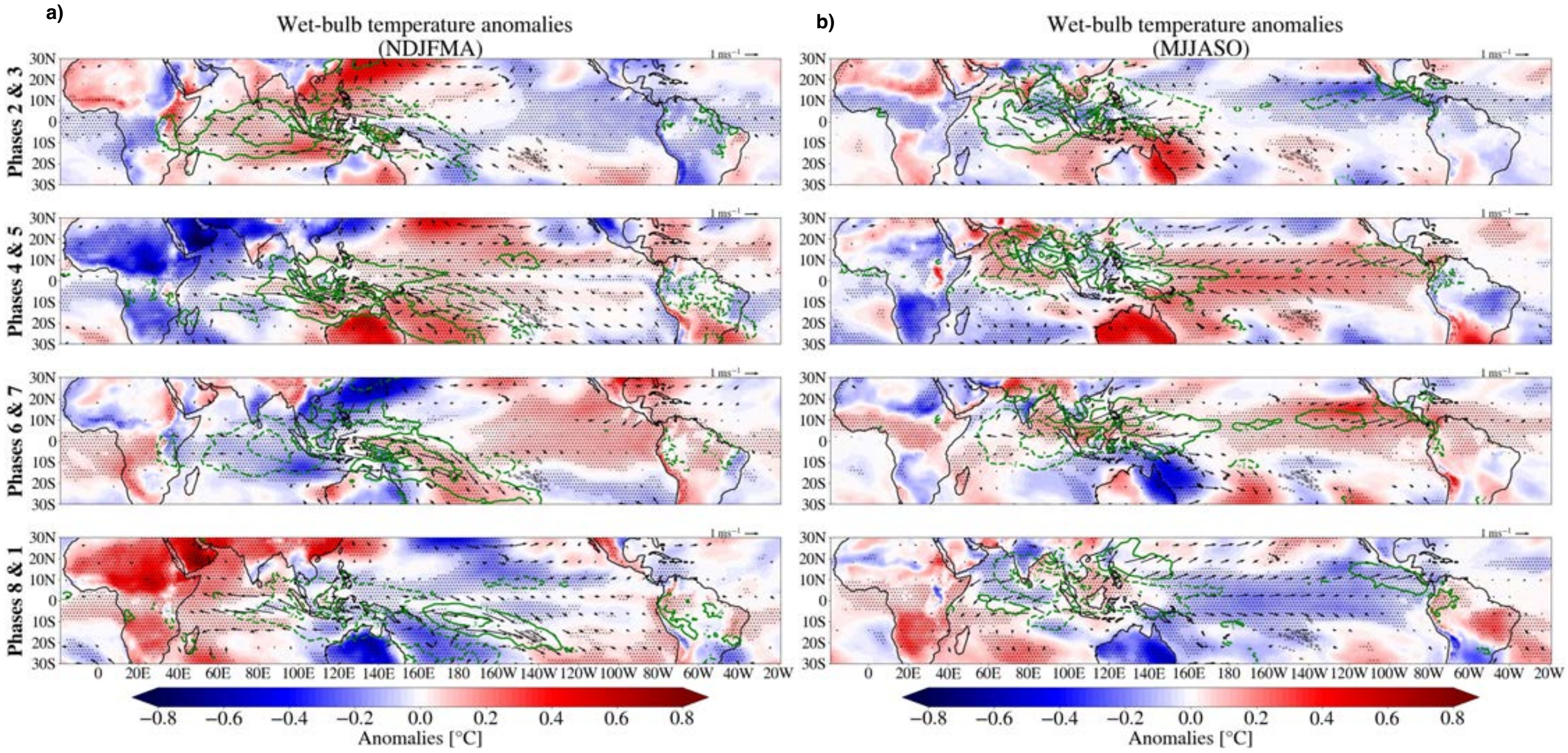

Fig. S5. Same as Fig. 1 but for Tw anomalies (°C).

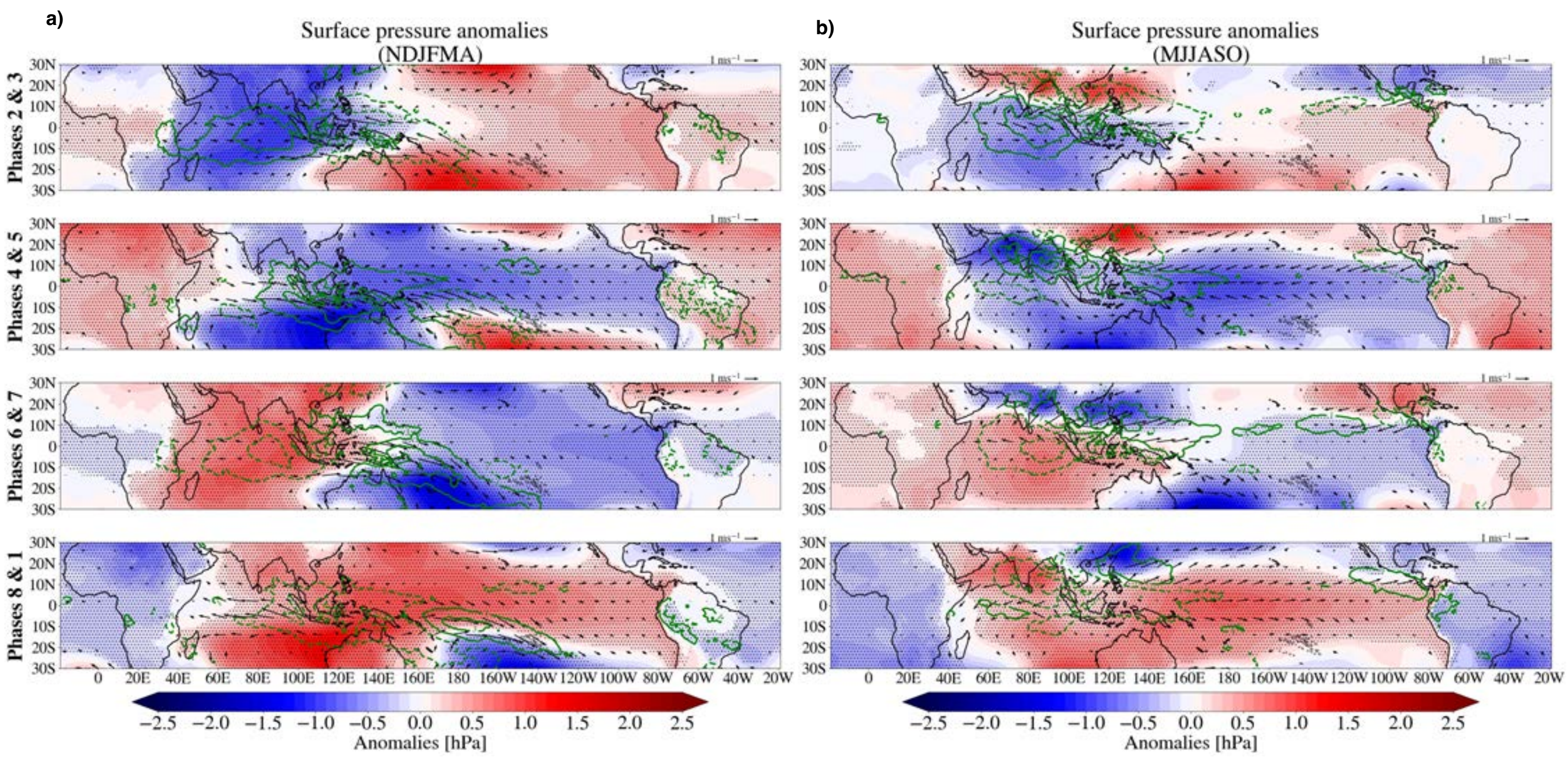

Fig. S6. Same Fig.1, for surface pressure anomalies (hPa).

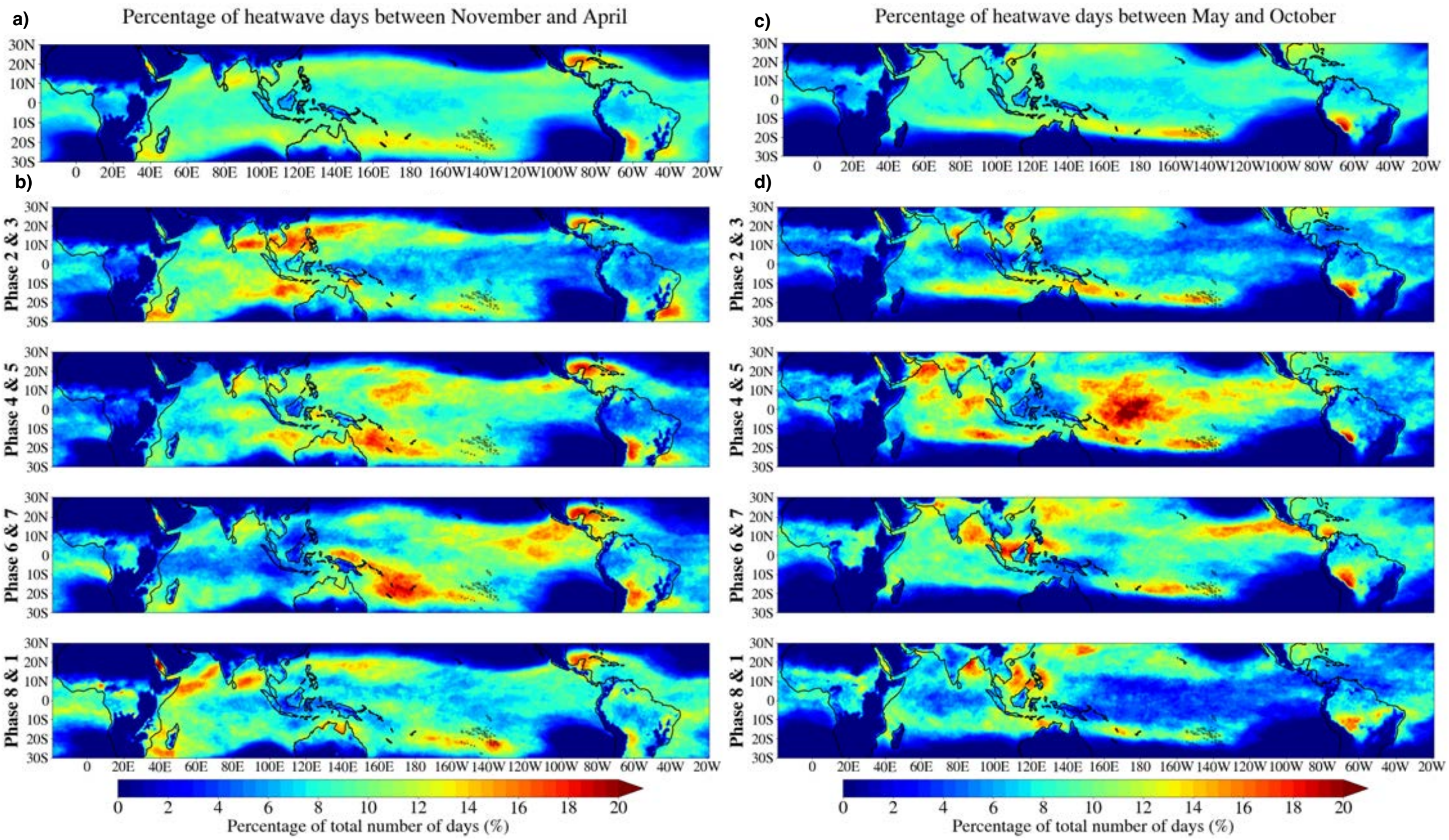

Fig. S7. Percentage of heatwave days per phase, meaning days where Tw exceeds the daily 90th percentile values (and remain above the minimum threshold of 21.9°C) (colours shading) for at least 3 consecutive days for the extended boreal winter (from November to April) on the left panels for a) the full 6-month long season *('Percentage$_{season_mean}$')*, b) each MJO phase pair *('Percentage$_{season_MJOphase}$')*. Right panels are the same for the extended boreal summer (from May to October).

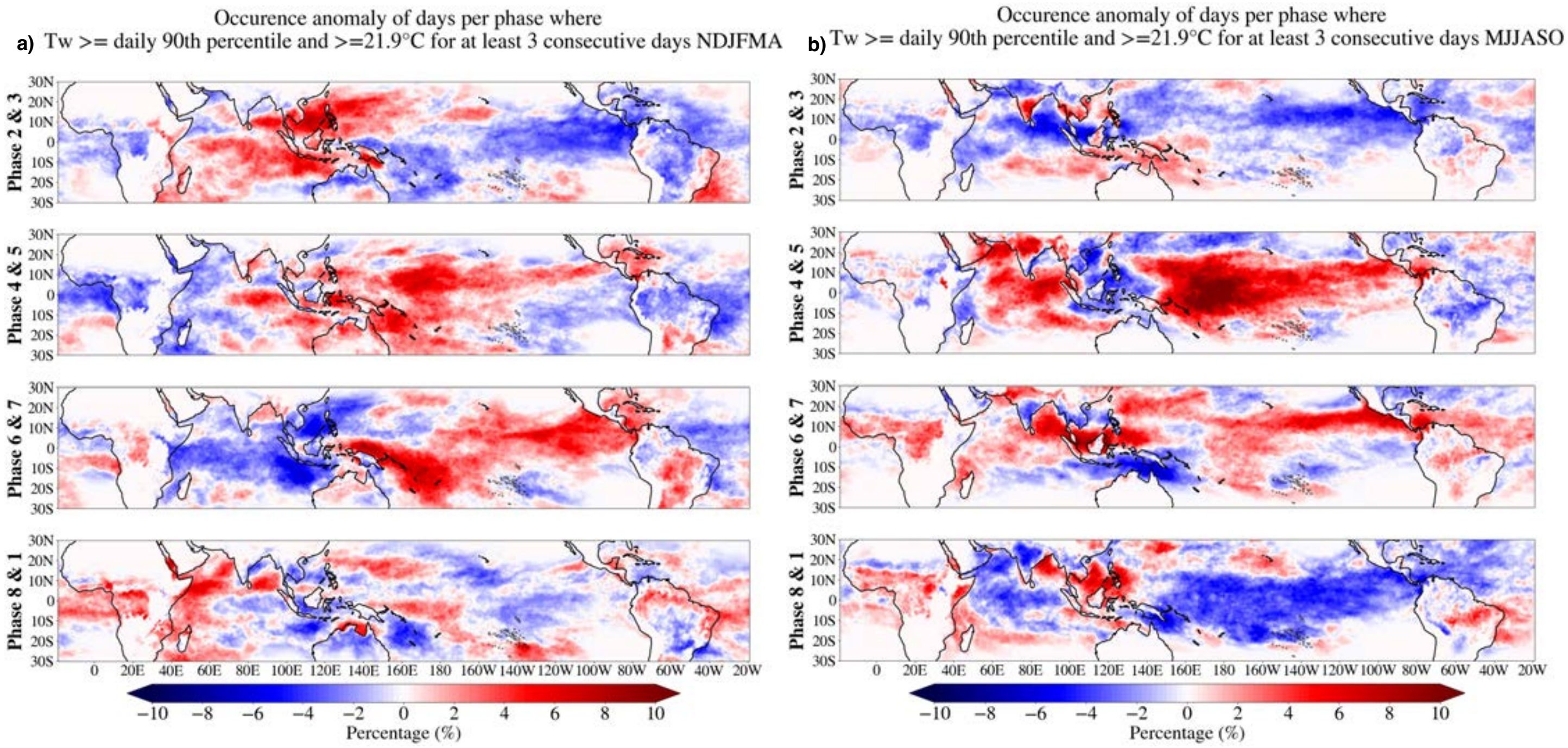

Fig. S8. The difference between the percentage of heatwave days during an MJO phase pair *('Percentage$_{season_MJOphase}$')* and the percentage of heatwave days for the season *('Percentage$_{season_mean}$')*. (a) NDJFMA and b) MJJASO). These maps provide an insight on the increase or decrease of the number of extreme humid heat days caused by the MJO at each grid-point for a given season.

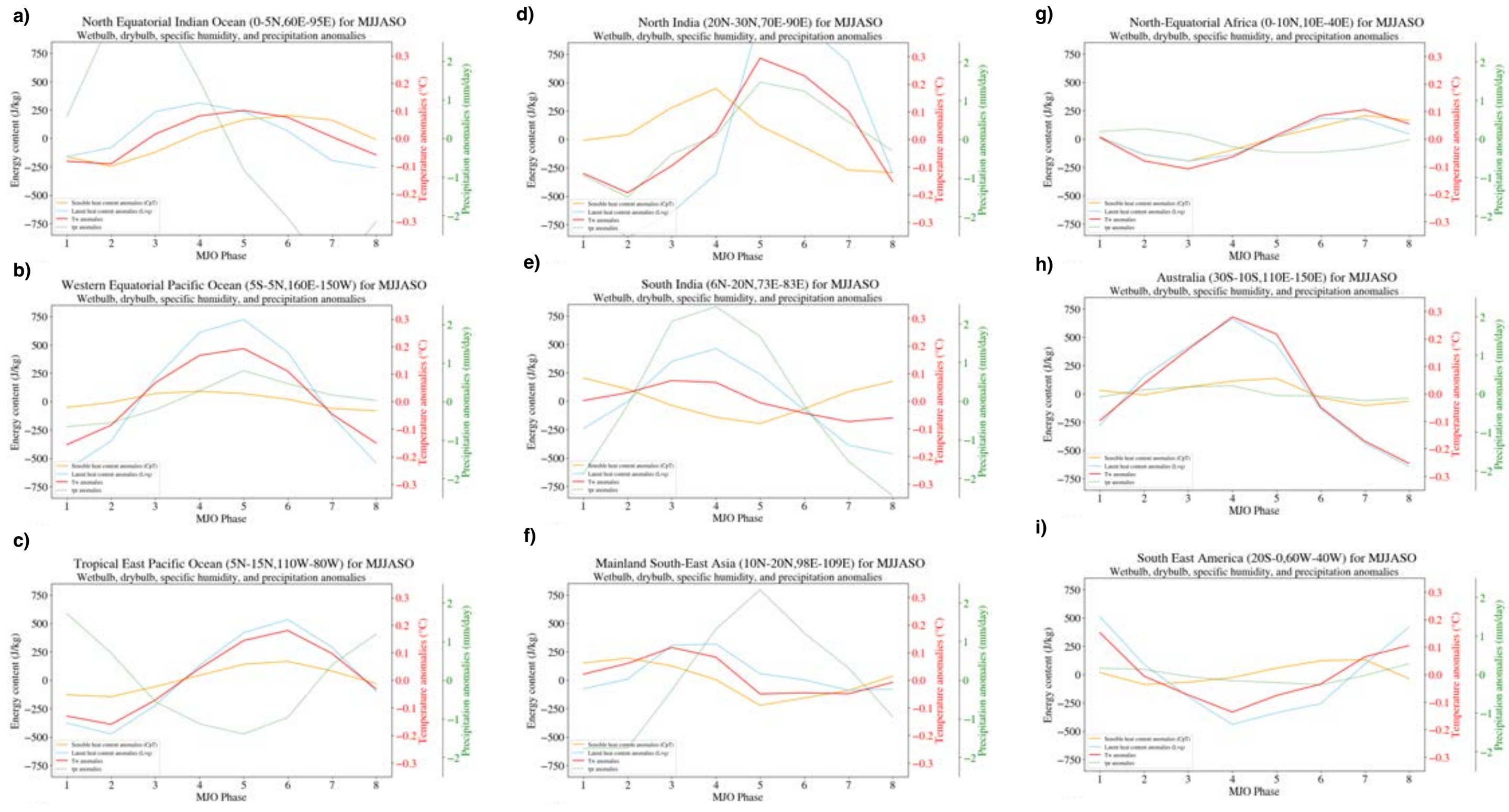

Fig. S9. As Fig. 8, but for MJJASO, for regions with strong MJO precipitation signals a-f) and little to no MJO precipitation g-i).

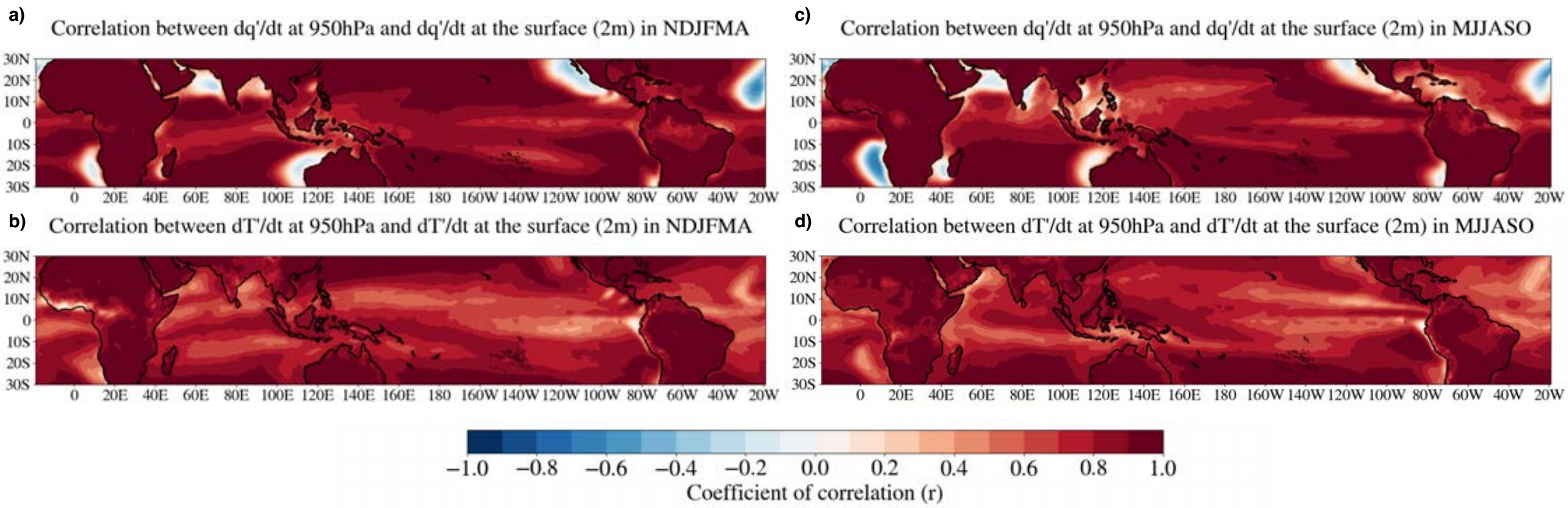

Fig. S10. Correlation (colour shading) between the tendency of specific humidity at 950 hPa and at 2 m (a,c), and between the tendency of temperature at 950 hPa and at 2 m (b,d). Results are shown for (a–b) NDJFMA and (c–d) MJJASO.

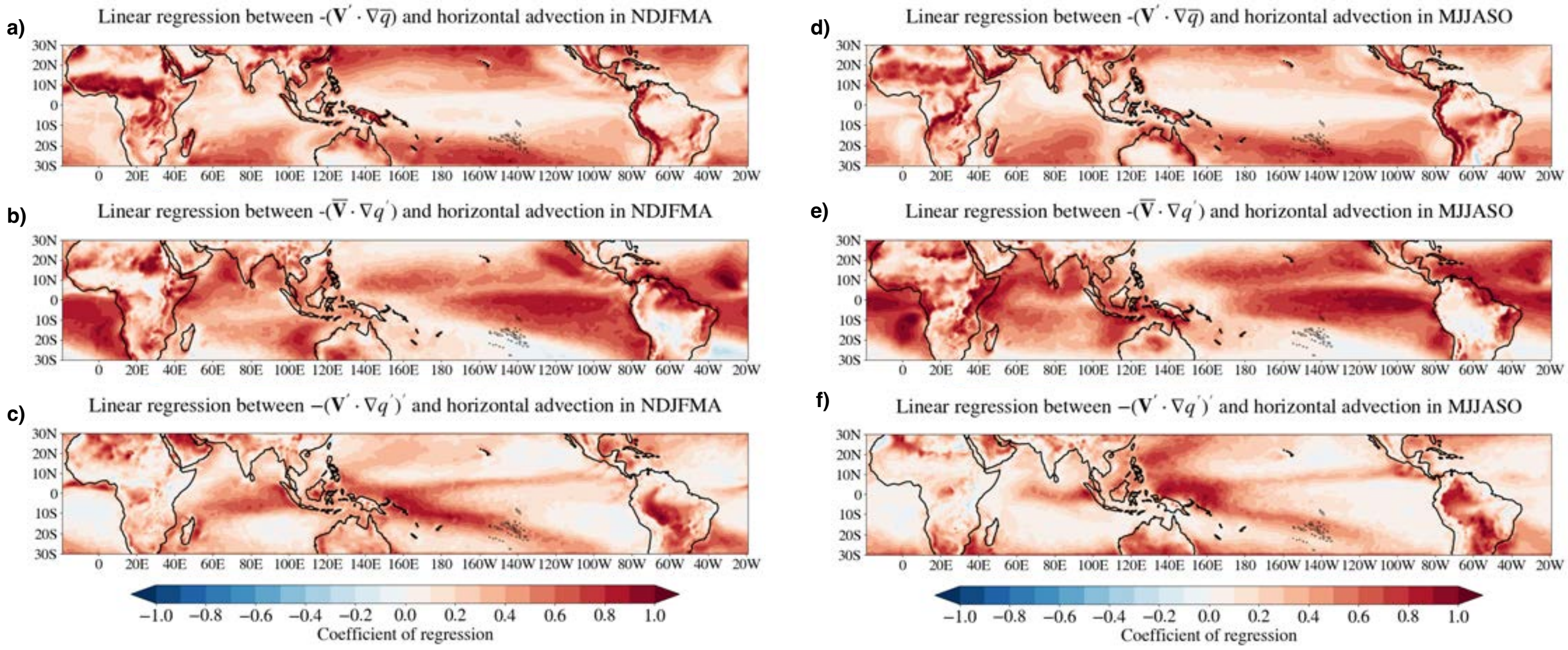

Fig. S11. Regression coefficient between horizontal advection of moisture anomalies at 950hPa and a) $\frac{\partial q\prime}{\partial t}$ at 950hPa, b) $\left(-\overrightarrow{\mathbf{V}'}.\nabla\overline{\mathrm{q}}\right)$ b) $\left(-\overrightarrow{\overline{\mathbf{V}}}.\nabla \mathrm{q}'\right)$ and c) $\left(-\overrightarrow{\mathbf{V}'}.\nabla \mathrm{q}'\right)$for NDJFMA. (e–h) as in (a–d) but for MJJASO for days when OMI >1.

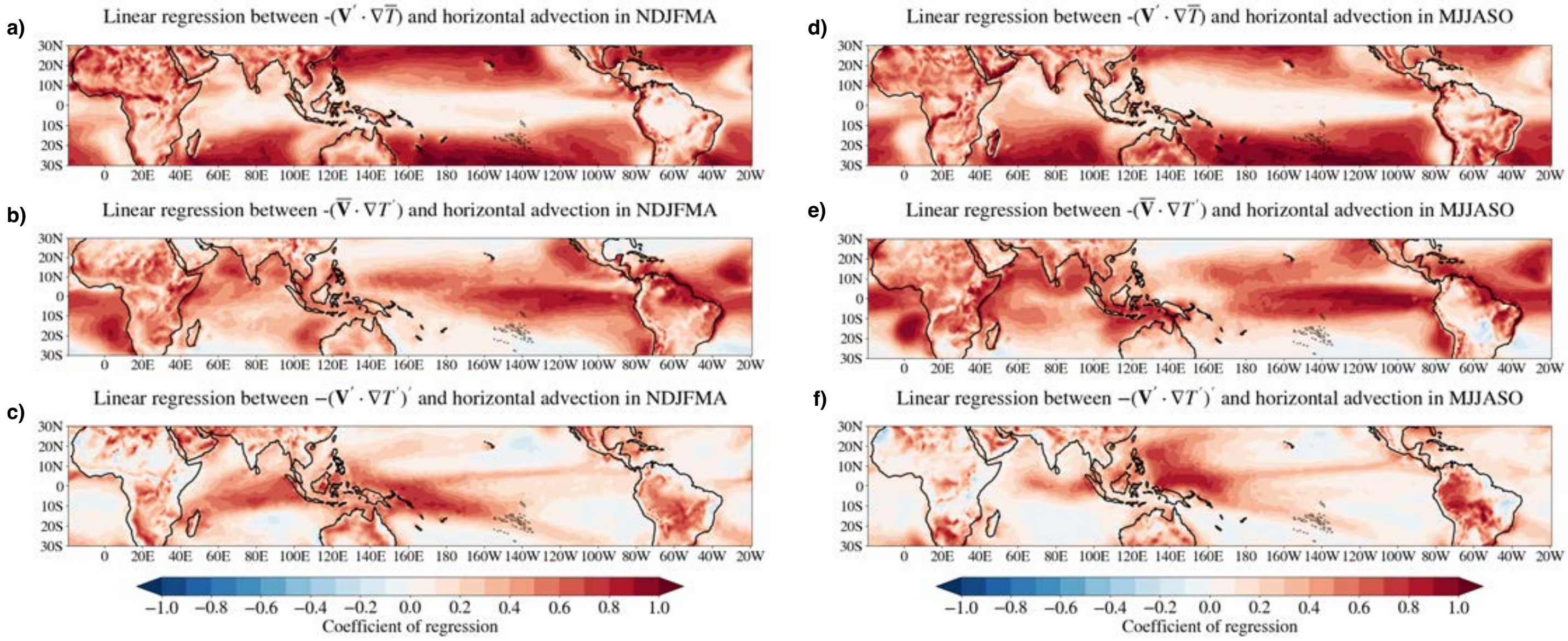

Fig. S12. Regression coefficient between horizontal advection of temperature anomalies at 950hPa and a) $\frac{\partial T\prime}{\partial t}$ at 950hPa, b) $\left(-\overrightarrow{\mathbf{V}'}.\nabla\overline{\mathrm{T}}\right)$, c) $\left(-\overrightarrow{\overline{\mathbf{V}}}.\nabla \mathrm{T}'\right)$, d) $\left(-\overrightarrow{\mathbf{V}'}.\nabla \mathrm{T}'\right)'$ for NDJFMA. (e–h) as in (a–d) but for MJJASO for days when OMI >1.

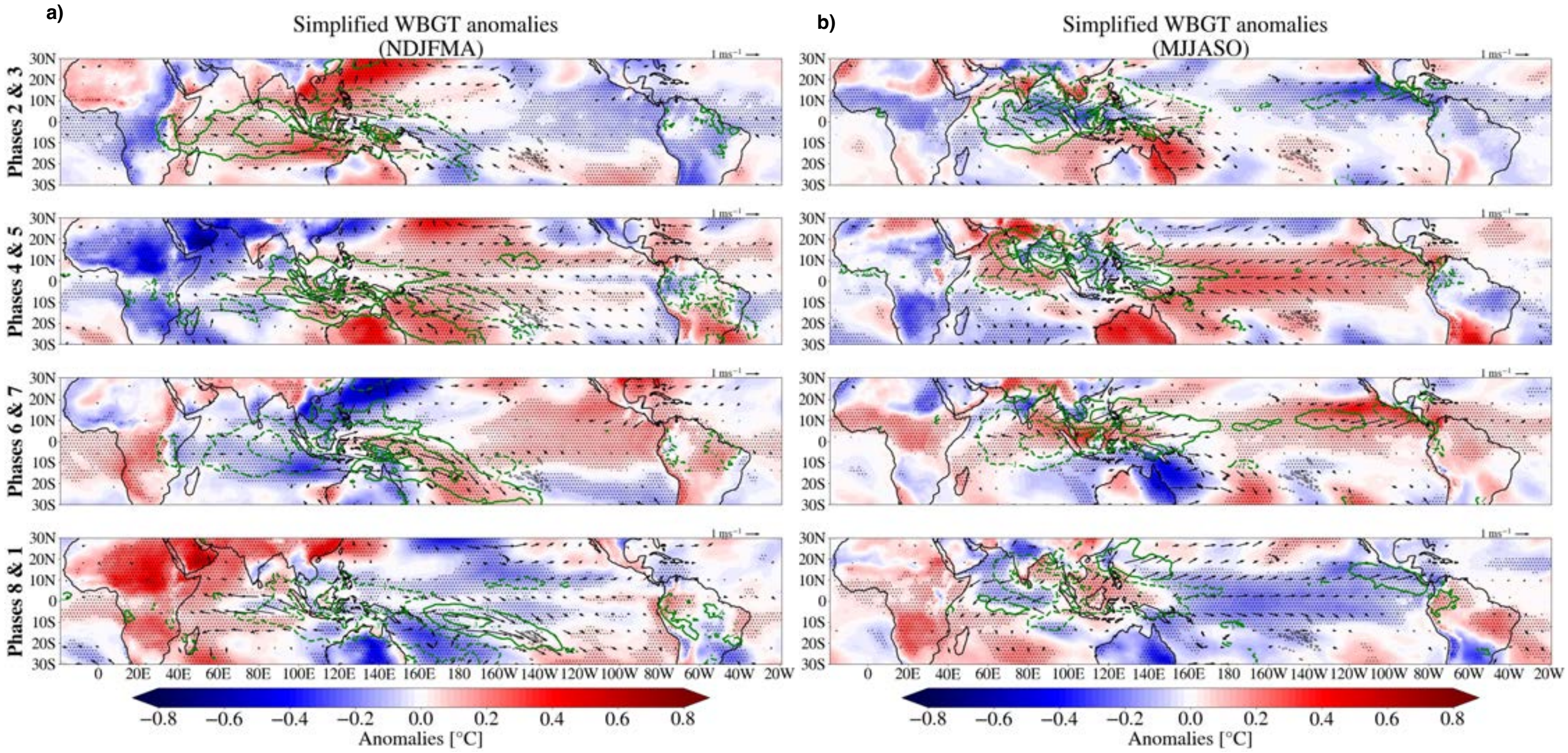

Fig. S13. Same as Fig. 1, for simplified wet bulb globe temperature (sWBGT) anomalies (°C).

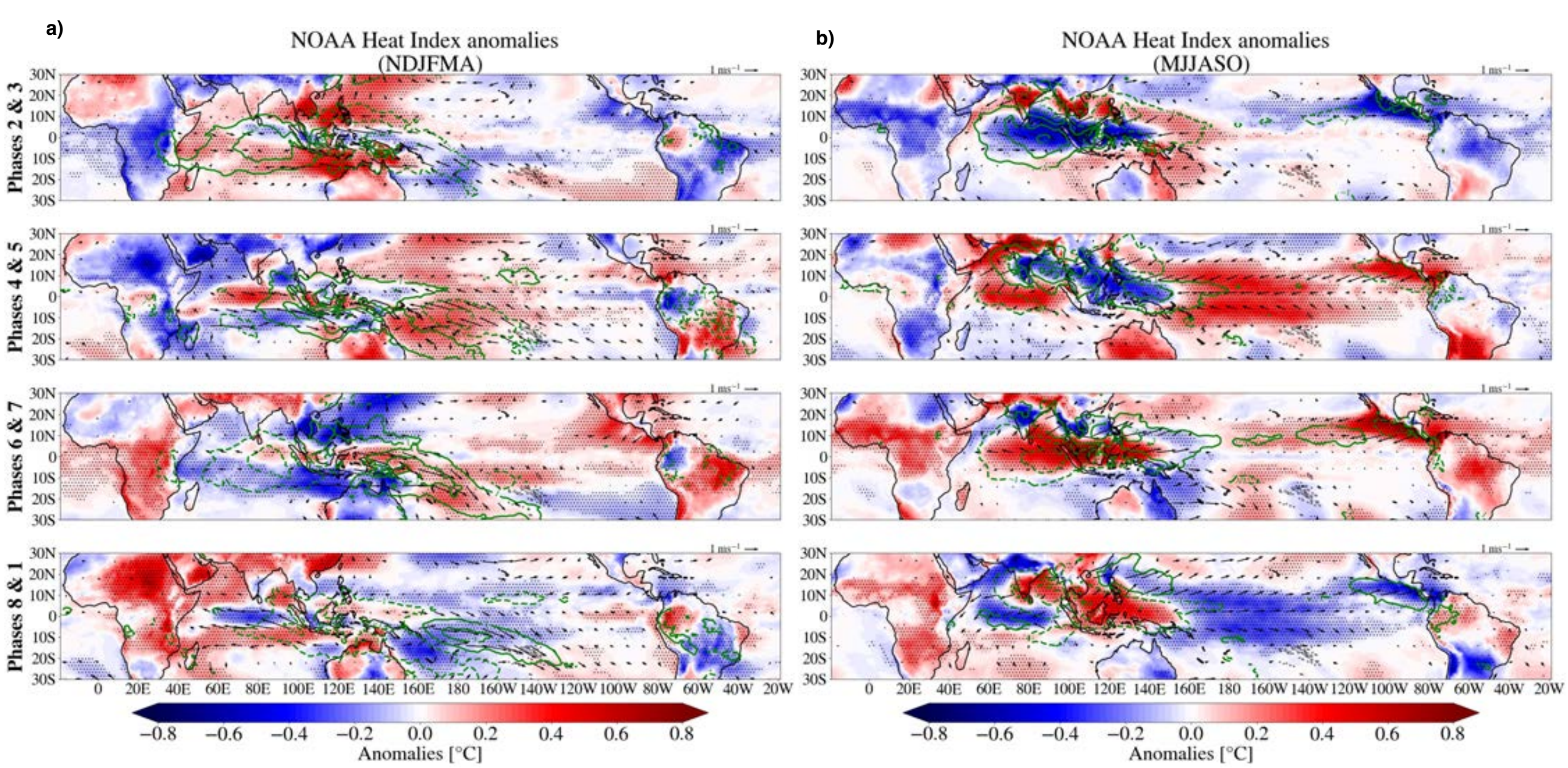

Fig. S14. Same as Fig. 1, for Heat Index (HI) anomalies (°C).

## 7. Diagnosis of the mechanisms behind regionally diverse relationship between MJO-related convection and humid heat

During the boreal winter, the MJO's direct influence on Tw temperatures via convection over northern Australia, is evident. During phases 8–1–2, when convection is suppressed, reduced cloud coverage and rainfall lead to increased dry-bulb temperatures. (Fig. 7, Fig. 5c). Specific humidity increases significantly during the build-up to and peak of convection (phases 3–4–5), coinciding with low-level westerly anomalies that draw moist tropical air eastward over the far north. (Fig. 6 and Fig. 5c). Concurrently, a low-level northerly wind anomaly carries warm, moist air southward into central Australia. This results in positive dry-bulb temperature and specific humidity anomalies over central Australia just before and during peak convective activity. These MJO-driven patterns align with established mechanisms described by Wheeler et al. (2009) and Marshall et al. (2022). During convective phases, dry-bulb temperatures over northern Australia are generally reduced due to increased cloud cover and rainfall (phases 4&5, Fig. 1 and Fig. 6). Despite this cooling, Tw anomalies remain positive. Strong positive specific humidity anomalies (Fig. 7) outweigh the reduction in dry-bulb temperature and absolute maximum Tw anomalies occur just prior to the convective peak (Fig. 5c). This timing represents a window in which specific humidity is increasing over the whole region, while dry-bulb temperatures in northern Australia have not yet been fully suppressed by convection. This demonstrates how humid heat can occur even when air temperature alone is not maximised.

In contrast, in regions with weaker convective signals, such as North Africa , South Tropical Africa (20S–5S, 10E–40E), the northwest Pacific (10N–30N, 120E–150E), and the northeast equatorial Pacific Ocean (0–10N, 120W–80W), the timing of the Tw anomalies is not necessarily linked to that of convection. Focusing on a region with weaker convective signals such as North Africa (10N–30N, 0–60E) in NDJFMA. During boreal winter and early spring, the MJO-driven circulation in the tropical Atlantic produces an eastward-propagating convective signal that extends across the entire basin and over Northern Africa (phases 8&1, Fig. 1). Although this convective signal is weaker than in the Indo-Pacific warm pool region, it amplifies low-level convergence, pulling moist air from the Atlantic and Gulf-of-Guinea northward and strengthens southerly/easterly flow that imports warm, moist air from the Indian Ocean into northern Africa in phases 8&1, directly increasing surface humidity and temperature (Zaitchik, 2017) and thereby leading to strong positive Tw anomalies (Fig. 5e). Over the subtropical northwest Pacific (10N–30N, 120E–150E), the MJO modulates surface Tw primarily through atmospheric teleconnections that modulate wind and thus horizontal advection (cf. subsection 3f). In phases 2& 3, enhanced convection over the eastern Indian Ocean and suppressed convection over the western Pacific force a circum-Pacific extratropical Rossby wave train from South Asia to Mexico (Moon et al., 2011). In the lower troposphere, this wavetrain leads to a large-scale anticyclonic anomaly that dominates the North Pacific (see surface pressure anomalies in Fig. S7 and Moon et al. (2011)), driving meridional southerly wind anomalies that advect warm, moist air from the tropics into the region which leads to positive Tw anomalies (Fig. 5g).

During the boreal summer, As the MJO propagates northeastward through phases 4& 5, southwesterly wind anomalies and increased wind speed enhance local evaporation in the North equatorial Indian Ocean (Fig. 9), coinciding with positive specific humidity anomalies. At the same time, dry-bulb temperatures begin to rise as OLR (precipitation) anomalies transition to positive (negative) values, marking the shift to suppressed convection. This transitional window, where moisture remains elevated and dry-bulb temperatures have not yet peaked, is where the highest-amplitude of warm Tw anomalies occur for this region (Fig. S10a). During the suppressed phase of the MJO over the western equatorial Pacific (phases 2&3), positive (negative) OLR (precipitation) anomalies indicate enhanced solar heating, which drives the initial rise in dry-bulb temperature in this region. Meanwhile, reinforced easterlies west of the convection center from phase 4 through phase 5, increase wind speed and enhance evaporation (Fig. 9). The propagation and initiation of the convective center via moisture convergence in the region, further contributes to moisture buildup during phases 4&5. Although evaporative cooling moderates surface temperatures, the concurrent increase in moisture dominates the Tw response, leading to peak humid heat anomalies during this period (Fig. S10b). In the northeast tropical Pacific, Tw is in phase with both dry-bulb temperature and specific humidity. During Phases 4&5 and 6&7, easterly wind anomalies and reduced wind speeds over the ocean surface promote surface warming through reduced latent heat flux (Fig. 7) and increased incoming solar radiation associated with diminished cloud cover (see positive (negative) OLR (precipitation) anomalies in Fig. 1). These easterlies also advect moist air from the coast of Mexico westward, contributing to positive specific humidity anomalies. Consequently, Tw peaks in phase 6, when both dry-bulb temperature and specific humidity reach their maximum (Fig. S10c). The subsequent transition to enhanced easterlies, increased wind speeds, and negative (positive) OLR (precipitation) anomalies reduces both dry-bulb temperature and specific humidity, driving negative Tw anomalies. In summary, whether through direct convective modulation or remote atmospheric teleconnections, the MJO creates distinct, phase-dependent environments that can significantly amplify humid heat stress.

## 8. Sensitivity test: comparing MJO impacts on heat stress using HI and sWBGT

Commonly used heat indices such as the simplified wet bulb globe temperature (sWBGT, Fig. S14) and the NOAA heat index (HI, Fig. S15) show significant anomalies associated with the MJO. The patterns and amplitude of the anomalies of the sWBGT are very similar to those of the Tw discussed in the main text. Whilst there are similarities in the global patterns between the HI and Tw, notably in the eastward propagation, distinct regional differences emerge. This is because the HI and Tw differ in their relative sensitivity to temperature versus humidity (cf. Fig. 6 and 7). Still, across both seasons, the HI anomalies are statistically significant over broad regions of the tropics and subtropics, with most areas being significantly affected by at least one MJO phase (but slightly less extended than Tw, as Ta significant anomalies are slightly less extended than specific humidity anomalies), even in regions typically not affected by MJO related convections. Moreover, the spatial structure of HI anomalies appears more strongly tied to the convection pattern than that of Tw, especially during MJJASO. HI anomalies tend to occur directly east and west of the MJO's convective centre, while pronounced negative anomalies are observed within the convective envelope, i.e. the dry-bulb temperature cooling driven by convection. These results highlight that, while all heat indices capture the MJO's influence to varying degrees, their differing sensitivities to temperature and humidity lead to distinct regional patterns and amplitudes, highlighting the importance of selecting appropriate metrics when assessing heat stress and its drivers.